\pdfoutput=1

\documentclass[twocolumn,aps,prl,superscriptaddress,showpacs,
amsfonts,amsmath]{revtex4}

\usepackage{amsthm}
\usepackage{braket}
\usepackage{graphicx}
\usepackage[colorlinks=true,citecolor=blue,urlcolor=blue]{hyperref}

\theoremstyle{definition}
\newtheorem{theorem}{Theorem}
\newcommand{\mean}[1]{{\langle{#1}\rangle}}
\def\endproof{\vrule height6pt width6pt depth0pt}

\begin{document}


\title{Necessary and sufficient condition for quantum state-independent contextuality}


\author{Ad\'an~Cabello}
 \email{adan@us.es}
 \affiliation{Departamento de F\'{\i}sica Aplicada II, Universidad de
 Sevilla, E-41012 Sevilla, Spain}

\author{Matthias Kleinmann}
 \email{matthias.kleinmann@uni-siegen.de}
 \affiliation{Department of Theoretical Physics, University of the Basque
 Country UPV/EHU, P.O.\ Box 644, E-48080 Bilbao, Spain}

\author{Costantino Budroni}
 \email{costantino.budroni@uni-siegen.de}
 \affiliation{Naturwissenschaftlich-Technische Fakult\"at, Universit\"at
 Siegen, Walter-Flex-Str.\ 3, D-57068 Siegen, Germany}


\date{\today}


\begin{abstract}
We solve the problem of whether a set of quantum tests reveals
state-independent contextuality and use this result to identify the simplest
set of the minimal dimension. We also show that identifying state-independent
contextuality graphs [R.\ Ramanathan and P.\ Horodecki, Phys.\ Rev.\ Lett.\
\textbf{112}, 040404 (2014)] is not sufficient for revealing state-independent
contextuality.
\end{abstract}



\maketitle


{\em Introduction.---}Contextuality, i.e., that the result of a measurement
does not reveal a preexisting value that is independent of the set of
comeasurable measurements jointly realized (i.e., the context of the
measurement), is one of the most striking features of quantum theory and has
been recently identified as a critical resource for quantum computing
\cite{HWVE14, Raussendorf13, DGBR14}. The earliest manifestation of
contextuality in quantum theory is the Kochen-Specker theorem \cite{Specker60,
KS67}, which states that, if the dimension $d$ of the quantum system is greater
than {2}, there exists a finite set of elementary tests (represented by
rank-one projectors in quantum theory) such that a value $1$ or $0$
(representing true or false, respectively) cannot be assigned to each of them
respecting that: (i) result $1$ cannot be assigned to two mutually exclusive
tests (represented in quantum theory by mutually orthogonal projectors), and
(ii) result $1$ must be assigned to exactly one of $d$ mutually exclusive
tests. Sets of elementary tests in which this assignment is impossible are
called {\em Kochen-Specker sets} \cite{PMMM05}.

Assumptions (i) and (ii) are not needed for detecting contextuality. It can be revealed by the violation of correlation inequalities satisfied by any model with noncontextual results. These inequalities are called {\em noncontextuality (NC) inequalities} \cite{SBKTP09}. Bell inequalities \cite{Bell64} are a special case of them.

Remarkably, there are NC inequalities which are violated by any quantum state
for a {\em fixed} set of measurements \cite{Cabello08}. A NC inequality with
this property is called a {\em state-independent NC (SI-NC) inequality},
whereas a set of elementary tests which can be used for such a
state-independent violation is called a {\em state-independent contextuality
(SIC) set}.

Every Kochen-Specker set is a SIC set \cite{BBCP09,YT14}, but there are SIC
sets that are not Kochen-Specker sets \cite{YO12, BBC12}. This observation,
together with the experimental implementation of SIC sets for testing SI-NC
inequalities \cite{KZGKGCBR09, ARBC09, MRCL10, ZWDCLHYD12, DHANBSC13,
CEGSXLC14} and the emergence of applications of SIC sets (e.g.,
device-independent secure communication \cite{HHHHPB10}, local
contextuality-based nonlocality \cite{Cabello10}, Bell inequalities revealing
full nonlocality \cite{AGACVMC12}, state-independent quantum dimension
witnessing \cite{GBCKL14}, and state-independent hardware certification
\cite{CAEGCXL14}) stimulated the interest in the problem of identifying SIC
sets.

In some cases, one can guess that a given set of elementary quantum tests is a
SIC set. Then, to prove it, it is sufficient to construct a SI-NC inequality
violated by these tests. For example, the set of elementary quantum tests
associated with the Peres-Mermin square \cite{Peres90, Mermin90} violates a SI-NC
inequality \cite{Cabello08}; therefore, it is a SIC set. However, in general,
one cannot follow this strategy and it is convenient to adopt a more general
point of view and consider not a specific set of elementary quantum tests, but
all sets of elementary quantum tests with a given exclusivity graph. In this
graph, vertices correspond to tests and edges occur when two tests are mutually
exclusive. Since elementary tests are represented by rank-one projectors and
two of them are mutually exclusive if and only if the corresponding projectors
are orthogonal, the exclusivity graph is equivalent to the orthogonality graph
of the corresponding projectors. This approach using graphs has been very
successful in investigating the general properties of quantum contextuality
\cite{CSW14, AFLS12} and the separation between quantum theory and other
hypothetical
theories \cite{Cabello13, FSABCLA13, Yan13, Sainz14, Cabello14}. An open question is when, for
a
given orthogonality graph, there exists a realization of the graph which is a
SIC set. Unfortunately, it has been notoriously difficult to answer this
question \cite{Cabello11}. The aim of this Letter is to provide a versatile
tool that allows one to approach this problem.

Recently, Ramanathan and Horodecki (RH) \cite{RH14} have presented a solution
to a relaxation of the problem of identifying SIC sets, namely of identifying
``SIC graphs.'' That is, whether a given graph admits, for any given state, a
realization as a set of projectors (with orthogonality relations corresponding
to edges in the graph) such that the correlations of such projectors on that
state violate some NC inequality.
This definition fits neither with the definition of a SIC set above nor with most
of the previous literature (cf.\ Refs.~\cite{Cabello08, BBCP09, YO12, BBC12,
Cabello10, AGACVMC12, GBCKL14, CAEGCXL14, Cabello11}). As far as we know, the
only work where a similar definition has been used is Ref.~\cite{KK12}.
Moreover, the definition of a SIC set in Ref.~\cite{RH14} is not
state independent on an operational level. The issue is that, according to
this definition, the realization of a SIC graph may depend on the state; the
set of measurements that violate the NC inequality may be different for
different initial states. Therefore, the definition is not state independent
on an operational level. To make an analogy, adopting a similar definition one
will reach the conclusion that a pentagon is a ``SIC graph for pure
states'' since any pure state will violate the
Klyachko-Can-Binicio\u{g}lu-Shumovsky NC inequality \cite{KCBS08} for some five
rank-one projectors whose orthogonality graph is a pentagon. In contrast, the
problem of identifying SIC sets not only has a long tradition (cf.\
Refs.~\cite{PMMM05, YO12, BBC12}), but also an immediate experimental
translation (cf.\ Refs.~\cite{ZWDCLHYD12, DHANBSC13, CEGSXLC14,CAEGCXL14}).

To prove that the result in Ref.~\cite{RH14} does not solve the problem of
identifying SIC sets, we begin by showing that there exists a SIC graph for
which no realization violates a NC inequality for every quantum state
(Theorem~\ref{sic-is-not-sic}). After that, we present a solution to the
problem of identifying SIC sets (Theorem~\ref{thm:hb}). Finally, we use it to
prove a conjecture formulated by Yu and Oh in Ref.~\cite{YO12} on the simplest
SIC set in $d=3$ (Theorem~\ref{yoconj}).

From graph theory we will use the notions of the chromatic number and the fractional chromatic number of a graph (cf.\ Ref.~\cite{SU11}). Given a graph $G$, i.e., a set
of vertices and the edges connecting them, a coloring of
the graph is an assignment of colors to vertices such that
vertices connected by an edge are
associated with different colors. The chromatic number $\chi(G)$
is the minimum number of colors needed. Similarly, the fractional
chromatic number $\chi_f(G)$ is the minimum of $\frac{a}{b}$ such
that vertices have $b$ associated colors, out of $a$ colors,
where vertices connected by an edge have associated disjoint sets of colors.
$\chi_f(G)$ can be computed as a linear program.

{\em Results.---}The operational state {\em dependence} of a SIC graph as
defined in Ref.~\cite{RH14} is apparent in the following theorem.


\begin{theorem}\label{sic-is-not-sic}
There exists a SIC graph for which no realization is a SIC set.
\end{theorem}


{\em Proof.---}%
In Ref.~\cite{RH14} it is proven that a necessary and sufficient
condition for a graph $G$ with a $[d, r]$-realization (i.e., a realization in
dimension $d$ by means of rank-$r$ projectors) to be a SIC graph is that the
fractional chromatic number $\chi_f(G)$ is strictly larger than $d/r$.

However, consider the {13}-vertex graph of Yu and Oh \cite{YO12},
$G_\mathrm{YO}$. This graph has a $[3, 1]$-realization and its fractional
chromatic number is $\chi_f(G_\mathrm{YO})=\frac{35}{11}$. Now consider the
{14}-vertex graph $G_{\mathrm{YO}+1}$ constructed by adding one vertex to
$G_\mathrm{YO}$ and linking this new vertex with the {13} vertices of
$G_\mathrm{YO}$. Clearly, this graph has a $[4, 1]$-realization and
$\chi_f(G_{\mathrm{YO}+1})= \frac{35}{11}+1 > 4$. It is true that, for any
state in $d=4$, there is a realization which violates a NC inequality. However,
whatever the realization, when the system is in the eigenstate corresponding to
the new vertex, there is an obvious noncontextual assignment of results, namely, one to
the $14$th projector and zero to all others.
\hfill\endproof


Now we will address the problem of identifying SIC sets. We first recall a
result from Ref.~\cite{KBLGC12} that helps us to identify sets of (not
necessarily rank-one) projectors for which there is a SI-NC inequality.


\begin{theorem}\label{sic-general}
A set of observables $\set{A_1, \dotsc, A_n}$ with spectra
$\{\sigma(A_1),\ldots,\sigma(A_n)\}$,
and contexts $C$ (i.e., the set of sets of comeasurable
observables) violates the SI-NC inequality
\begin{equation}\label{eq:sicth2}
\sum_{c\in C} \lambda_c \mean{\prod_{k\in c} A_k} \le \eta
\end{equation}
with $0\le \eta < 1$ and real coefficients $\lambda_c$, if and only if
\begin{equation}
\sum_{c\in C} \lambda_c \prod_{k\in c} a_k \le \eta
\text{ for all } a \text{ and }
\sum_{c\in C} \lambda_c \prod_{k\in c} A_k \geq \openone,
\end{equation}
where the entries $a_k$ in $a=(a_1, \dotsc, a_n)$ assume any value from $\sigma(A_k)$.
\end{theorem}


Then, the necessary and sufficient condition for a set of rank-one projectors to constitute a SIC set is given by the following.


\begin{theorem}\label{thm:hb}
A set of rank-one projectors $S=\set{\Pi_1, \dotsc, \Pi_n}$ is a SIC set if and
only if there are non-negative numbers $w=(w_1, w_2, \dotsc)$ and a number $0\le
y<1$ such that
\begin{equation}\label{e:sdpth2}
 \sum_{j\in \mathcal I} w_j \le y \text{ for all $\mathcal I$ and }
 \sum_i w_i \Pi_i\ge \openone,
\end{equation}
where $\mathcal I$ is any set such that $i, j\in \mathcal I$ implies
$\Pi_i\Pi_j\ne 0$ (i.e., $\mathcal I$ is any independent set of the
orthogonality graph of $S$).

In particular, $w$ gives rise to the SI-NC inequality
\begin{equation}\label{e:ncgen}
 \sum_i w_i \mean{\Pi_i} - \sum_{i} w_i \sum_{j\in \mathcal N(i)}
\mean{\Pi_i\Pi_j}\le y,
\end{equation}
 where $\mathcal N(i)=\set{j| \Pi_i \Pi_j=0}$ is the orthogonality neighborhood
of $i$.
\end{theorem}


{\em Proof.---}%
For proving sufficiency, we will prove that, for a given $(y, w)$ satisfying
 conditions~\eqref{e:sdpth2}, with $0\leq y<1$, inequality~\eqref{e:ncgen} is a
 valid NC inequality and it is violated for every state.
For that, it is enough to realize that among the noncontextual assignments
 maximizing the left-hand side of inequality~\eqref{e:ncgen} are those that
 respect the orthogonality conditions; i.e., two orthogonal projectors could not both
 have been assigned the value $1$.
Respecting the orthogonality conditions precisely amounts to assign $1$ to the
 elements of a set $\mathcal{I}$ appearing in conditions~\eqref{e:sdpth2} and,
 hence, the bound $y$ holds for inequality~\eqref{e:ncgen}.
The proof goes as follows.
Let us consider orthogonal projectors $\Pi_i$ and $\Pi_j$ and any noncontextual
 assignment $p\in\{0, 1\}^n$ such that $p_i=1$ but $p_j=0$.
By changing the value of $p_j$, i.e., violating the orthogonality condition, we
 get an extra contribution $w_j$ from the first term and $-\sum_{k\in\mathcal
 N(j)} (w_j+w_k) p_k\le -w_j$ from the second term, decreasing the total value
 of the left-hand side of inequality~\eqref{e:ncgen}.
This proves that inequality~\eqref{e:ncgen} is a valid NC inequality.
By condition~\eqref{e:sdpth2}, it is violated by any quantum state.

For proving necessity, we show that if $\{\Pi_i\}$ give rise to a violation of
 a linear NC inequality for every state, then conditions~\eqref{e:sdpth2} are
 satisfied.
Let us assume, for some $(\lambda,\eta)$, that the following inequality is
 violated by any state
\begin{equation}\label{SICP}
\sum_{\mathcal C} {\lambda}_{\mathcal C} \mean{\prod_{k\in
 \mathcal C} \Pi_k }\leq {\eta},
\end{equation}
 where the sum is over all cliques $\mathcal C$ different from the empty set in
 the orthogonality graph of $S$, corresponding to all possible contexts, and
 $\lambda_\mathcal C$ are real numbers.
Notice that the use of a linear expression in inequality~\eqref{SICP} is not a
 restriction as it follows from the Hahn-Banach theorem (cf., e.g.,
 Ref.~\cite{Horn85}).
In fact, the set of quantum correlations for all states and the set of
 noncontextual correlations are (compact) convex sets, and hence the sets
 either intersect or they can be separated by a hyperplane, i.e., distinguished
 via a linear inequality.
Notice also that inequality~\eqref{SICP} contains all of the possible correlations
 that are jointly measurable; i.e., it includes all contexts $\mathcal{C}$,
 with a generic coefficient $\lambda$.

Since inequality~\eqref{SICP} holds, in particular, for all assignments
 respecting orthogonality, we have $\sum_{k\in \mathcal I} {\lambda}_{\set{k}}
 \leq \eta$ for any independent set $\mathcal I$.
At the same time, we assume a state-independent violation and hence, without
 loss of generality, $\sum_k \lambda_{\set{k}} \Pi_k \ge \openone$ and $\eta
 <1$.
[In general we have $\sum_k \lambda_{\set{k}}\Pi_k \ge \xi\openone$ and
 $\eta<\xi$.
But the assignment $p\equiv (0, 0, \dotsc)$ yields $0\le \eta< \xi$, which
 allows us to rescale $\lambda_\mathcal{C}\rightarrow \lambda_\mathcal{C}/\xi$
 and $\eta\rightarrow \eta/\xi$.]
Eventually, we identify $w_i=\max\set{0, \lambda_{\set{i}}}$ and $y=\eta$.
Indeed, inequality~\eqref{SICP} has to hold for any assignment $p=(p_1, \dotsc,
 p_n)$ respecting orthogonality and having $p_k=0$ for all
 $\lambda_{\set{k}}<0$.
This way, the condition in Eq.~\eqref{e:sdpth2} is obeyed by that
 identification.
\hfill \endproof
\\
We mention that the condition in Theorem~\ref{sic-general} as well as that in
Theorem~\ref{thm:hb} can be verified by means of a semidefinite program.
Semidefinite programs are a class of optimization problems that can be solved
numerically with a certificate of optimality \cite{SDP}.


At this point, it is interesting to point out the relation between
Theorem~\ref{thm:hb} and the results in Ref.~\cite{RH14}. According to
Ref.~\cite{RH14}, to conclude that a graph of orthogonality is a SIC graph, it
is sufficient to check the expectation value of $\sum_j w_i \Pi_i$ on the
maximally mixed state $\rho=\openone/d$. Assuming rank-one projectors, we can
substitute the condition $\sum_i w_i \Pi_i\ge \openone$ with $\frac{1}{d}\sum_i
w_i \ge 1$, yielding RH's result. In fact, the condition in Eq.~\eqref{e:sdpth2} can be
formulated in terms of the existence of a solution greater than $d$ for the
linear program
\begin{equation}\begin{split}\label{e:lpproj2}
\text{maximize: } &\sum_i w_i \\
\text{subject to: }&
 \sum_{j\in \mathcal I} w_j \le 1 \text{ for all $\mathcal I$,} \\
 & w_i\ge 0 \text{ for all $i$.}
\end{split}\end{equation}
Every $(w, y)$ obeying Eq.~\eqref{e:sdpth2} with $y<1$ can be used to achieve
$\sum_i w_i>d$ by rescaling all the weights by $1/y$.
The linear program in Eq.~\eqref{e:lpproj2} is the dual problem
of the fractional chromatic number $\chi_f(G)$ of the orthogonality graph $G$
(also known as the fractional clique number, cf.\ Ref.~\cite{SU11}); hence, both yield
the same optimal value.

Together with the fact that the chromatic number $\chi(G)$ is lower bounded by the fractional
chromatic number $\chi_f(G)$ \cite{SU11}, we have the following.


\begin{theorem}\label{nc}
Necessary conditions for a set of rank-one projectors in dimension $d$ to be a SIC set are that for the orthogonality graph $G$, (i) $\chi_f(G)>d$ and (ii) $\chi(G)>d$.
\end{theorem}
Condition (i) is also a direct consequence of the results in Ref.~\cite{RH14},
where it was demonstrated in addition that, in general, condition (ii) is
strictly weaker than condition (i). However, condition (ii) has the advantage
of being solvable exactly by simple integer arithmetic, while condition (i) is
the solution to a linear program.


The minimal dimension in which SIC sets exist is $d=3$ \cite{KS67}. Therefore,
identifying the smallest SIC set in $d=3$ is a problem of fundamental
importance. Using the previous results we can prove a conjecture from
Ref.~\cite{YO12}.


\begin{theorem}\label{yoconj}
In dimension $d=3$, there exists no SIC set with less than {13} projectors. The
set provided by Yu and Oh in Ref.~\cite{YO12} is therefore the simplest for
$d=3$.
\end{theorem}


{\em Proof.---}%
The orthogonality graph of a SIC set has to obey at least the
following necessary conditions: (a) that the graph has a $[3,
1]$-representation, and (b) that the graph has a fractional chromatic number
greater than {3}.

{From} condition (a) it follows that the graph must be square free, because for
a projector represented by a vertex of the square, the other two connected to
it must be in the orthogonal plane, and the fourth is orthogonal to both, so it
must be the same as the first.

The first step is to generate all nonisomorphic, i.e., not obtained via a relabeling, square-free connected graphs with {12} or fewer vertices and then calculate their chromatic
number. It is sufficient to consider connected graphs since for a disconnected graph the chromatic number is the largest chromatic number of its connected components.

\begin{figure}
{\includegraphics[width=\linewidth]{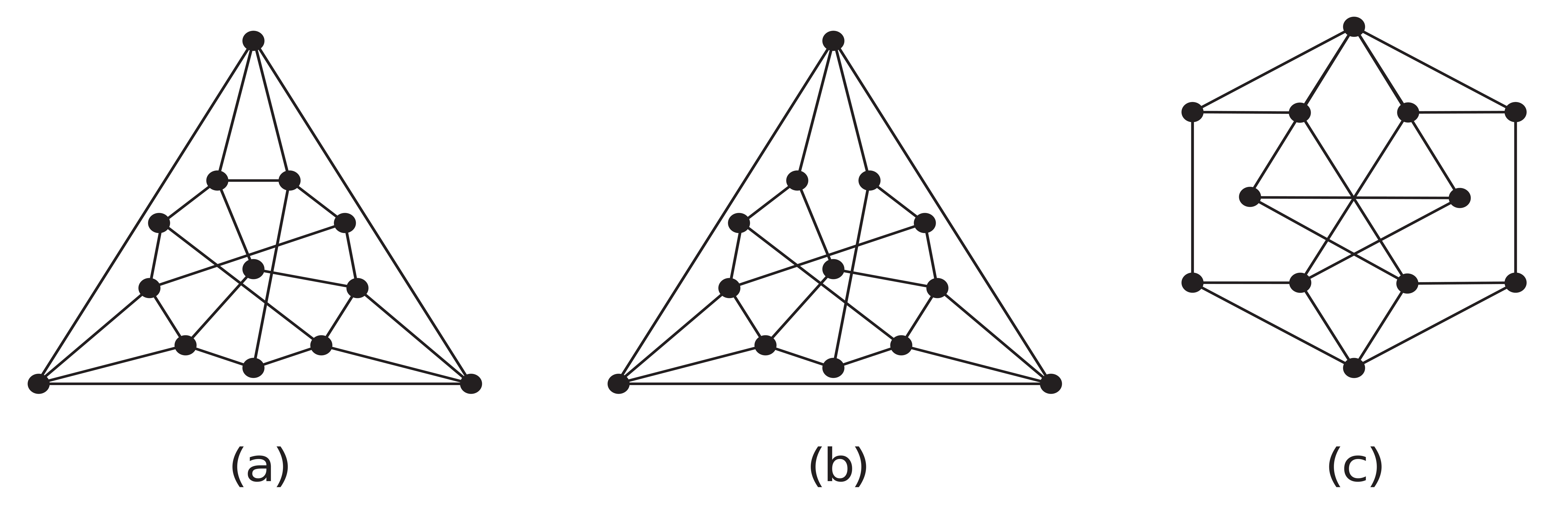}}
\caption{\label{Fig01} (a) Yu-Oh graph $G_\mathrm{YO}$, (b) $G_\mathrm{YO}$ minus one
edge. (c) The only square-free connected {12}-vertex graph with chromatic
number $\chi(G)>3$.}
\end{figure}

For this, we use the utility \texttt{geng} from the software package \texttt{nauty}
{v2.5r9} \cite{nauty}, and we find $143\,129$ graphs with such properties.
Among them, there is only one graph $G$ with $\chi(G)>3$, which is depicted in Fig.~\ref{Fig01}(c). By solving the linear program in Eq.~\eqref{e:lpproj2} with exact arithmetic~\cite{Fukuda}, one finds that its fractional chromatic number is $\chi_f(G)=3$.
\hfill \endproof

One can go further and ask whether there are other SIC graphs with {13}
vertices aside from the Yu-Oh graph $G_\mathrm{YO}$, depicted in Fig.~\ref{Fig01}(a).
There are in total eight square-free graphs with {13} vertices and $\chi(G)>3$
\footnote{In the \texttt{graph6} encoding
(\url{http://cs.anu.edu.au/\~ebdm/data/formats.html}), they are:
``\texttt{L{?}AEB{?}oDDIQSUS}'',
``\texttt{L{?}AEB{?}oFDHISPS}'',
``\texttt{L{?}ABA\_oo\_iREJa}'',
``\texttt{L{?}ABAagF{@}bWgHc}'',
``\texttt{L{?}ABEagE{`}gH{`}{`}c}'' (which is $G_\mathrm{YO}$ minus one edge),
``\texttt{L{?}AB{?}vOLDPHa{`}o}'' (which is $G_\mathrm{YO}$),
``\texttt{L{?}BDA\_gEREHcac}, and
``\texttt{L{?}{`}D{@}bCUCbDgWc}''.},
and out of these eight graphs, only three have
$\chi_f(G)>3$ \footnote{In the \texttt{graph6} encoding,
``\texttt{L{?}ABEagE{`}gH{`}{`}c}'', having $\chi_f=19/6$,
``\texttt{L{?}AB{?}vOLDPHa{`}o}'', having $\chi_f=35/11$, and
``\texttt{L{?}{`}D{@}bCUCbDgWc}'', having $\chi_f=13/4$.
}. Two of them are depicted in Fig.~\ref{Fig01}, (a) $G_\mathrm{YO}$ and
(b) $G_\mathrm{YO}$ minus one edge, together with the {12}-vertex graph (c), which is
a common induced subgraph of all remaining {13}-vertex graphs with $\chi(G)>3$.

The existence of a representation in dimension $d=3$ for a given orthogonality graph can be written as a minimization of a polynomial function. In fact, the scalar product of two complex vectors can be written as a polynomial with vector entries as variables; hence, orthogonality conditions correspond to its zeros. A numerical search was not able to find a solution for such graphs in $d=3$.


{\em Conclusion.---}We have started arguing that the definition of ``state-independent contextuality scenario'' used in Ref.~\cite{RH14} is inconsistent with almost all of the previous literature on the topic and is not
state independent on an operational level because the realization of the scenario depends on the state. Then we have shown that the criterion proposed in Ref.~\cite{RH14} does not solve the problem of whether or not a set of
quantum tests reveals state-independent contextuality in the sense defined in
most of the literature, including all experimental implementations and
applications. Then we have presented a solution to this problem and explained
the connection between this solution and the results in Ref.~\cite{RH14}.
Finally, we have used our result to prove that the Yu-Oh set is the simplest
set of elementary quantum tests revealing state-independent contextuality in
dimension three. Our results clarify the structure of state-independent
contextuality and---as we demonstrated on an example---enable the systematic
investigation of state-independent contextuality sets.


\begin{acknowledgments}
The authors thank P. Horodecki, J. R. Portillo, R. Ramanathan, and S. Severini
for the useful conversations. This work was supported by Project No.\ FIS2011-29400
(MINECO, Spain) with FEDER funds, the FQXi large grant project ``The Nature of
Information in Sequential Quantum Measurements,'' and by the European Union (ERC Starting
Grant No.\ GEDENTQOPT). Theorems~\ref{nc} and~\ref{yoconj} incorporate results that
appeared in an unpublished article \cite{Cabello11}.
\end{acknowledgments}



\end{document}